\documentclass[aps,prb,reprint,superscriptaddress,amsmath,amssymb,floatfix,longbibliography]{revtex4-2}

\usepackage[utf8]{inputenc}
\usepackage[T1]{fontenc}
\usepackage{graphicx}
\usepackage{bm}
\usepackage{mathtools}
\usepackage{amsfonts}
\usepackage{xcolor}
\usepackage{xr-hyper}
\usepackage[colorlinks=true,allcolors=blue!60!black]{hyperref}

\newcommand{\Z}{\mathbb{Z}}
\newcommand{\C}{\mathbb{C}}
\newcommand{\im}{\operatorname{Im}}

\newcommand{\Lz}{J_{z}}
\newcommand{\ee}{\mathrm{e}}
\newcommand{\ii}{\mathrm{i}}
\DeclareMathOperator{\sgn}{sgn}

\begin{document}

\title{Integer-Graded Reciprocal Skin Effect from Internal Rotation}

\author{Kin Hung Fung}
\email[Corresponding author: ]{funguiuc@gmail.com}
\affiliation{Department of Physics, The Hong Kong University of Science and Technology, Hong Kong, China}

\author{C.~T.~Chan}
\affiliation{Department of Physics, The Hong Kong University of Science and Technology, Hong Kong, China}
\affiliation{Institute for Advanced Study, The Hong Kong University of Science and Technology, Hong Kong, China}

\date{\today}

\begin{abstract}
A prominent class of reciprocal non-Hermitian skin effects has a $\Z_{2}$ topological classification,
characterized by a parity invariant protected by reciprocity. We show that conserved
internal rotation refines this classification into an unbounded integer-graded hierarchy. Every angular momentum channel is an exactly solvable
asymmetric chain with its own integer point-gap winding, while conjugate channels accumulate at
opposite edges so that reciprocity is preserved globally. For conserved channels,
the classification contains one integer per conjugate pair, with the number of independent integers
determined by the internal multiplet. Half-integer spin realizes the symplectic class
AII$^{\dagger}$, whose $\Z_{2}$ index is recovered as the parity of the grading. Conjugate-channel
mixing destroys the grading and leaves a critical, scale-free skin effect. A cyclic three-site
lattice and a passive array of lossy resonators provide minimal realizations.
\end{abstract}

\maketitle

\section{Introduction}
\label{sec:intro}

The non-Hermitian skin effect (NHSE)---the accumulation of an extensive number of bulk eigenstates at a boundary under open boundary conditions (OBC)---has become a major topic in the study of open and nonconservative systems, and has been
realized in photonic and acoustic lattices, mechanical and active metamaterials, electrical
circuits, and cold-atom and open electronic systems~\cite{Lee2016,Yao2018,Kunst2018,Gong2018,MartinezAlvarez2018,FoaTorres2020,Bergholtz2021,Ashida2020,Okuma2023,Lin2023,Zhang2022}.
In one dimension its origin is point-gap topology: a nonzero winding of the spectrum about a
reference energy controls the boundary accumulation and the associated semi-infinite boundary
index~\cite{Zhang2020,Okuma2020,Borgnia2020}, and the generalized Brillouin zone (GBZ) restores
the bulk--boundary correspondence between periodic boundary conditions (PBC) and OBC~\cite{Yao2018,Yokomizo2019,Yang2020}. A standard example is the
Hatano--Nelson chain~\cite{Hatano1996,HatanoNelson1997}, whose right and left hopping amplitudes are unequal.

\begin{figure}[!t]
\centering
\includegraphics[width=\columnwidth]{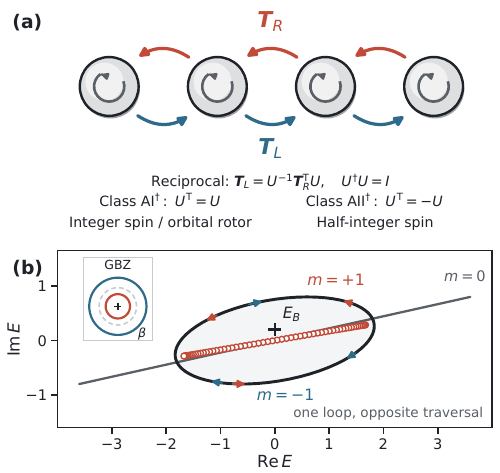}
\caption{\label{fig:main}%
The reciprocal rotor chain and its graded skin effect. (a) Identical cells carry a compact rotor,
and the inter-cell coupling of Eq.~\eqref{eq:TRTL} is reciprocal as a matrix
[Eq.~\eqref{eq:recip}]. The paired arrows denote the reciprocal coupling matrices $T_R$ and $T_L$;
$T_R$ maps cell $n+1$ to $n$, and $T_L$ maps cell $n$ to $n+1$. The unitary reciprocity matrices satisfy $U^{\mathsf T}=U$ in class AI$^{\dagger}$
and $U^{\mathsf T}=-U$ in class AII$^{\dagger}$. (b) For the AI$^{\dagger}$ example at $\alpha=2\pi/3$, the channels $m=\pm1$ trace the same
PBC loop but traverse it in \emph{opposite senses} (terracotta/teal), so their
windings about the common base point $E_{B}=0.2\ii$ (cross, shaded) are $\nu_{\pm1}=\pm1$, while
the $m=0$ locus degenerates to a zero-area segment through the origin and does not wind. The base
point stays clear of every channel spectrum by at least $0.195$ in units of $t_{0}$. Under OBC, each loop collapses onto an interior segment (points). Inset: the channel-resolved GBZs
are exact circles [Eq.~\eqref{eq:gbzradius}], with $m\equiv1,2\ (\mathrm{mod}\ 3)$ sitting on
reciprocal radii about the Bloch circle (dashed). Parameters $t_{0}=1$, $t_{1}=0.8+0.4\ii$.}
\end{figure}

This model, like most of its extensions, therefore starts from manifest nonreciprocity. A
broader question is whether boundary accumulation can instead emerge only after resolving an
internal symmetry, while the full coupling remains reciprocal. Although the underlying boundary
physics is different, the quantum spin Hall effect provides a useful symmetry-based comparison:
time-reversal symmetry protects helical edge states in a phase
characterized by a $\Z_{2}$ invariant~\cite{KaneMele2005,BHZ2006,HasanKane2010}. The
non-Hermitian setting admits a distinct boundary phenomenon: in a \emph{reciprocal} skin effect,
partners connected by reciprocity accumulate at opposite boundaries even though the full
Hamiltonian remains reciprocal. Such reciprocal skin effects have been reported in several
settings. Topolectrical circuits and acoustic metamaterials show reciprocal partners accumulating at opposite ends, including
projective-mirror and purely lossy realizations~\cite{Hofmann2020,Liu2025Z2acoustic,Wang2026passiveZ2}, while point-gap topology combined with
internal-symmetry constraints admits time-reversal-protected and bidirectional skin
effects~\cite{Okuma2020,Kawabata2020symplectic,Wang2024symmetry}. Orbital systems, models with spin-dependent gauge fields or non-Abelian couplings, and
ring-resonator and optical systems likewise exhibit internal-state-resolved directionality~\cite{Chen2025orbital,Zhang2025SU2,Han2026helical,Tang2026nonabelian,Xu2021ring,Takeda2025circulating}, and a separate route obtains an NHSE from spatial
noncommutativity, $[\hat{x},\hat{y}]=\ii\theta$~\cite{WeiKou2026}. Recent extensions have uncovered crystalline-symmetry-protected $\Z_{4}$ skin
topology, spin-resolved skin effects induced by gauge fields, general symmetry classifications, and
symmetry-fractionalized skin phenomena in interacting systems~\cite{Ishikawa2024glide,Sanahal2025spin,Shiozaki2026intrinsic,Ekman2026fractionalized}. These developments reveal increasingly rich symmetry-resolved skin phenomena, but do not provide a general microscopic mechanism for organizing reciprocal skin modes into a hierarchy with distinct point-gap windings.

We show that conserved internal rotation provides such a mechanism. Every cell of a chain carries
an internal rotation, and neighboring cells are joined by two parallel coupling paths: a direct
path that leaves the internal state unchanged and a phase-carrying path. The full coupling matrices
obey the reciprocity relation, so the chain is strictly reciprocal. Angular momentum is nevertheless
conserved, so each fixed-$m$ channel sees the same two complex amplitudes weighted by conjugate
phase factors, $\ee^{-\ii m\alpha}$ and $\ee^{+\ii m\alpha}$, in the two inter-cell blocks.
These two scalar sums generally have different magnitudes, so each channel behaves as a
nonreciprocal Hatano--Nelson chain even though the full coupling matrix is reciprocal. Reciprocity maps a
channel onto its partner of opposite angular momentum and reverses the bias, so the partner
accumulates at the other end. A single Kramers pair with odd channel winding realizes the reciprocal $\Z_{2}$ skin effect; an AI$^{\dagger}$ pair instead needs the conserved channel label. Larger internal multiplets support an integer-graded hierarchy.

The paper is organized as follows. Sections~\ref{sec:model}--\ref{sec:c3} introduce the reciprocal
rotor chain, derive its angular-momentum channels, and present the cyclic three-site realization.
Sections~\ref{sec:gbz}--\ref{sec:spin} describe the graded localization lengths, rotation-resolved topology,
and half-integer-spin case. Sections~\ref{sec:passive} and~\ref{sec:numerics} present a passive realization
and numerical demonstrations. Sections~\ref{sec:discussion}--\ref{sec:conclusion} discuss the physical
interpretation and stability and conclude with an outlook.
Appendices~\ref{app:range}--\ref{app:related} give the finite-range realization, passive
implementation, numerical checks, and a comparison with related gradings. The Supplemental
Material~\cite{SM} gives the explicit real-space Hamiltonian of the minimal cyclic three-site model, detailed half-integer-spin and
symmetry-class analyses, transport derivations, and analyses of absorption and nonuniform perturbations.

\section{Reciprocal rotor chain}
\label{sec:model}

Before resolving the internal states, we state the symmetry requirements on the full Hamiltonian.
It conserves an internal-rotation generator $J_z$,
$[H,J_z]=0$, and reciprocity reverses its eigenvalue. We denote by $U$ the unitary matrix acting within
each cell that implements the reversal, $U J_z U^{-1}=-J_z$. The full lattice Hamiltonian
then
satisfies the transpose-type reciprocity condition

\begin{equation}
H^{\mathsf{T}}=UHU^{-1}.
\label{eq:recip}
\end{equation}
Here $U$ acts identically in every cell. For the physical reciprocity operation considered here, we distinguish
$U^{\mathsf T}=U$, corresponding to class AI$^{\dagger}$ for orbital rotors,
cyclic lattices, and integer spin, from $U^{\mathsf T}=-U$, corresponding to
class AII$^{\dagger}$ for genuine half-integer spin~\cite{Kawabata2019}.
In the former case a suitable local unitary basis makes $U$ the identity, so
Eq.~\eqref{eq:recip} reduces to $H^{\mathsf T}=H$, as in the site basis of the cyclic lattice
in Sec.~\ref{sec:c3}. In the latter case the physical reciprocity matrix is antisymmetric
and cannot be transformed to the identity, giving the symplectic, Kramers-paired structure
discussed in Sec.~\ref{sec:spin}.
For a translationally invariant chain, the same condition
becomes $H^{\mathsf{T}}(k)=U H(-k)U^{-1}$ in Bloch space~\cite{Kawabata2019,Wang2024symmetry}.
A minimal nearest-neighbor realization places a compact rotor in every cell of a one-dimensional
lattice [Fig.~\ref{fig:main}(a)]. With cells
$n=1,\dots,N$ and internal amplitude vectors $\psi_n$, the Hamiltonian acts as
\begin{equation}
(H\psi)_{n}=h_{0}(J_z)\,\psi_{n}+T_{R}\,\psi_{n+1}+T_{L}\,\psi_{n-1},
\label{eq:H}
\end{equation}
with $\psi_{0}=\psi_{N+1}=0$ for the open chain and
\begin{equation}
T_{R}=t_{0}\,I+t_{1}\,\ee^{-\ii\alpha J_z},\qquad
T_{L}=t_{0}\,I+t_{1}\,\ee^{+\ii\alpha J_z}.
\label{eq:TRTL}
\end{equation}
Here $t_{0}$ and $t_{1}$ are complex amplitudes of two parallel inter-cell paths: a direct path
that preserves the internal state and a rotation-carrying path. The rotation-carrying path introduces opposite phase factors
$\ee^{-\ii m\alpha}$ and $\ee^{+\ii m\alpha}$ in the two hopping directions for each angular-momentum component $m$. A nonzero $\alpha$ is engineered by a finite relative
rotation of this coupling in the internal space: for an orbital rotor, an azimuthal displacement
$\alpha$ of the coupling point produces the internal rotation $\ee^{-\ii\alpha J_z}$; for a cyclic $C_n$ multiplet, shifting the
coupling by one internal site gives $\alpha=2\pi/n$ (written out site by site in
Supplemental Material~\cite{SM}, Sec.~\ref{ext:sm:realspace}). The reverse-direction block carries
$\ee^{+\ii\alpha J_z}$, so this finite rotation does not itself break reciprocity. At $\alpha=0$,
$T_R=T_L$ and the channel skin bias vanishes. The on-site term is an even function,
$h_{0}(-J_z)=h_{0}(J_z)$; a constant $\varepsilon_{0}$ suffices and is used throughout.
All terms are functions of $J_z$, so the internal rotation is conserved identically. Conjugation by
$U$ sends $\ee^{-\ii\alpha J_z}$ to $\ee^{+\ii\alpha J_z}$ and hence exchanges $T_R$ and $T_L$,
while leaving $h_0$ unchanged. Because these internal matrices are symmetric in the $J_z$ basis,
transposition of the full lattice Hamiltonian exchanges the two
inter-cell blocks and gives
Eq.~\eqref{eq:recip}. Thus the chain is reciprocal before any channel is selected; the skin bias,
if present, must emerge only after symmetry resolution.

\section{Angular-momentum channels and directional hopping}
\label{sec:channels}

Because $J_z$ is conserved,
its eigenvalue $m$ labels independent \emph{channels}. The label may run over $m\in\mathbb Z$
for an orbital-type internal space or over $m=-j,\dots,j$ for a spin-$j$ multiplet, integer or
half-integer. For the latter, $U=\ee^{-\ii\pi J_y}$ realizes $m\leftrightarrow -m$ and obeys
$U^{\mathsf T}=(-1)^{2j}U$. Within a channel of eigenvalue $m$, the rotation operator
$\ee^{-\ii\alpha J_z}$ reduces to the scalar phase factor $\ee^{-\ii m\alpha}$ and
$h_0(J_z)$ to $\varepsilon_m$, leaving a Hatano--Nelson-type single-band
chain~\cite{Hatano1996,HatanoNelson1997}. Its two inter-cell hopping amplitudes are
\begin{equation}
r_{m}=t_{0}+t_{1}\,\ee^{-\ii m\alpha},\qquad
l_{m}=t_{0}+t_{1}\,\ee^{+\ii m\alpha},
\label{eq:rl}
\end{equation}
and its Bloch Hamiltonian is
$H_{m}(k)=\varepsilon_{m}+r_{m}\ee^{\ii ka}+l_{m}\ee^{-\ii ka}$, where $a$ is the lattice constant
(Supplemental Material~\cite{SM}, Secs.~S2--S3).

Equation~\eqref{eq:rl} directly shows the origin of the channel asymmetry. In the $T_{R}$ block, the phase-carrying path enters relative to the direct path with the phase factor
$\ee^{-\ii m\alpha}$; in the $T_{L}$ block, it enters with the conjugate phase factor
$\ee^{+\ii m\alpha}$. The same two complex
amplitudes, combined with conjugate
phase
factors, generally have different magnitudes, so the channel is biased even
though the full Hamiltonian remains reciprocal. The size of the bias follows in
closed form,

\begin{equation}
g_{m}\equiv|r_{m}|^{2}-|l_{m}|^{2}=4\,\im(t_{0}^{*}t_{1})\,\sin(m\alpha),
\label{eq:generator}
\end{equation}
a single real number that we call the generator.

Equation~\eqref{eq:generator} has three immediate consequences. The trivial channel is unbiased, $g_{0}=0$, because
the
rotation acts trivially there. Conjugate channels are biased oppositely, $g_{-m}=-g_{m}$, so they
localize at opposite ends.
And the bias is proportional to $\im(t_{0}^{*}t_{1})$, the area of the parallelogram spanned by the two amplitudes in the complex plane, so it is nonzero only when the two paths are \emph{non-collinear}. Thus neither path on its own breaks left--right symmetry; the asymmetry arises only from their interference, and only after the internal phase assigns conjugate factors to the two inter-cell blocks. Both ingredients are needed---the amplitudes must be out of phase, $\arg t_{1}\neq\arg t_{0}\,(\mathrm{mod}\,\pi)$,
and the rotation phase must satisfy $\sin(m\alpha)\neq0$.

The generator also has an energy-dissipation interpretation in the passive realization of
Appendix~\ref{app:passive}. The coherent path, $t_1=t$, carries the internal-rotation phase,
whereas the dissipative path, $t_0=-\ii c$, is rotation-blind, with $t,c>0$~\cite{Metelmann2015}.
For energy-normalized resonator amplitudes, each shared lossy element is driven by the sum of
the fields in neighboring cells. In-phase contributions dissipate more energy, whereas equal
and opposite contributions do not excite that bond-loss channel. The least-damped Bloch wave,
$ka=\pi$, therefore has a velocity set by the coherent coupling,
$v_m^*=-2ta\sin(m\alpha)=-ag_m/(2c)$ [Eq.~\eqref{ext:sm:vstar}]. A single-cell excitation loses its
more strongly damped components first, so its normalized center of mass approaches motion at
$v_m^*$ before reaching a boundary. Thus $g_m>0$ gives leftward drift and left-end skin
accumulation, whereas $g_m<0$ gives the opposite direction. Conjugate channels have opposite
biases, although the dissipative coupling remains reciprocal. The normalized drift and the
skin eigenmode envelope are distinct consequences of this interference, not amplification
[SM Sec.~\ref{ext:sm:transport}].

\emph{Scalar limit of the present mechanism.} A one-component reciprocal bond has no nontrivial
rotation eigenvalue that distinguishes its two paths. Within
Eqs.~\eqref{eq:H}--\eqref{eq:TRTL} a scalar internal space carries only the trivial character:
the rotation reduces to unity, the two paths merge into the single amplitude $t_{0}+t_{1}$ in both
directions, and the generator vanishes identically. The mechanism therefore needs a multimode
internal space carrying a nontrivial rotation representation, together with non-collinear
amplitudes. This does not exclude scalar skin effects produced by other means---explicit
nonreciprocity, longer-range winding, Floquet driving, or higher-dimensional
geometry~\cite{Hatano1996,Zhang2020,Zhang2022higherD}. A minimal lattice realization now shows that
the required internal phase can be implemented in an ordinary reciprocal network.

\section{Concrete realization: the cyclic three-site lattice}
\label{sec:c3}

\begin{figure}[!t]
\centering
\includegraphics[width=\columnwidth]{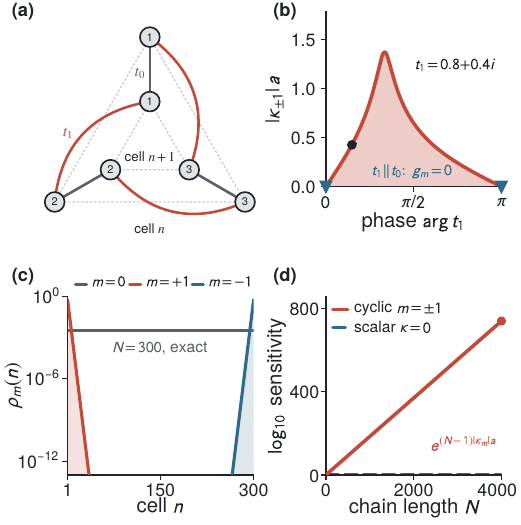}
\caption{\label{fig:cyclic3}%
Concrete cyclic three-site ($C_{3}$) realization. Panels (b)--(d) use the closed-form solution, so
every curve is exact at arbitrary $N$ and no diagonalization is involved. (a) End-on view of two adjacent cells, $n$ (outer triangle) and $n+1$ (inner triangle), showing direct bonds $t_0$ (gray) and cyclically shifted bonds $t_1$ (red). All bonds are reciprocal. Dashed triangles only outline the cells; sites within each cell are uncoupled. (b) Skin rate of the $m=\pm1$ channels versus the coupling phase $\arg t_{1}$ at fixed
$|t_{1}|$. It vanishes where the two couplings become collinear, $\arg t_{1}=0,\pi$, since the
generator of Eq.~\eqref{eq:generator} vanishes there and every channel is gauge-equivalent to a reciprocal chain, i.e.\
NHSE-free within this mechanism; the marker shows the working point $t_{1}=0.8+0.4\ii$. (c) Exact
channel-resolved skin profile at $N=300$: $m=0$ extended, $m=\pm1$ at opposite edges at equal rate, with the
envelope $\rho_{m}(n)\propto|\beta_{m}|^{2n}$ of Eqs.~\eqref{eq:gbzradius}
and~\eqref{eq:mirror}. The profile is intensive, i.e.\ independent of $N$, because the rate
$\kappa_{m}$ is. (d) Condition number of the similarity transformation that removes the skin,
which measures how sensitive the spectrum is to perturbations. It grows exponentially with $N$ for
every localized channel and stays flat for a scalar chain. Parameters $t_{0}=1$,
$t_{1}=0.8+0.4\ii$.}
\end{figure}

The rotor need not be abstract. A minimal and manifestly reciprocal realization puts three sites---coupled resonators, waveguides,
or circuit nodes---in each cell, and takes the internal rotation to be the $C_{3}$ cyclic shift
$S$, the $3\times3$ permutation matrix that sends site $j$ to site $j{+}1\ (\mathrm{mod}\ 3)$. The
coupling matrices of Eq.~\eqref{eq:TRTL} then become $t_{0}I+t_{1}S$ and $t_{0}I+t_{1}S^{\dagger}$ [Fig.~\ref{fig:cyclic3}(a)]. In the physical site basis, these $3\times3$ matrices are
\begin{equation*}
\begin{aligned}
S&=\begin{pmatrix}0&0&1\\1&0&0\\0&1&0\end{pmatrix},\\[2pt]
T_R&=t_0I+t_1S=\begin{pmatrix}t_0&0&t_1\\t_1&t_0&0\\0&t_1&t_0\end{pmatrix},\\[2pt]
T_L&=t_0I+t_1S^{\mathsf T}=\begin{pmatrix}t_0&t_1&0\\0&t_0&t_1\\t_1&0&t_0\end{pmatrix}=T_R^{\mathsf T}.
\end{aligned}
\end{equation*}
In terms of bonds, $t_{0}$ joins sites of equal index in neighboring cells, $t_{1}$ joins site
$j{+}1$ of cell $n$ to site $j$ of cell $n{+}1$, and sites of the same cell are not coupled;
Supplemental Material~\cite{SM}, Sec.~\ref{ext:sm:realspace} gives the Hamiltonian of two adjacent cells and a drawing of the
network. Since $S$ is a permutation, every
inter-site bond equals its reverse and $H$ is complex symmetric: reciprocity is manifest at the
level of the network, with no internal reversal needed.

The unitary $F_{jm}=\omega^{mj}/\sqrt{3}$, with $\omega=\ee^{2\pi\ii/3}$, diagonalizes the shift and
gives the three characters $m=0,\pm1$ at $\alpha=2\pi/3$ and the exact amplitudes of
 Eq.~\eqref{eq:rl} with the rotation phases replaced by $\omega^{\mp m}$. The outcome is the
minimal graded reciprocal skin effect: $m=0$ stays extended, while $m=\pm1$ accumulate at opposite
ends at equal rate [Fig.~\ref{fig:cyclic3}(c)]. The imbalance switches off when the two couplings
are made collinear [Fig.~\ref{fig:cyclic3}(b)], and the sensitivity of the spectrum to
perturbations grows exponentially with the length of the chain [Fig.~\ref{fig:cyclic3}(d)].
The character transformation $F$ is complex unitary rather than orthogonal, and satisfies
$F^{\mathsf{T}}F=U$ with $U$ the character reflection $m\leftrightarrow-m$. Thus the manifest
network symmetry $H=H^{\mathsf{T}}$ in the site basis is equivalent to the reciprocity
constraint~\eqref{eq:recip} in the character basis---no genuine angular momentum is required
to realize it.

The microscopic construction identifies the channel bias; its boundary consequence is determined by
the non-Bloch problem for each channel.

\section{Graded localization lengths}
\label{sec:gbz}

Under OBC, the channel spectrum and states are
governed not by $H_{m}(k)$ on the unit circle but by its continuation onto the GBZ~\cite{Yao2018,Yokomizo2019}.
After the substitution $\ee^{\ii ka}\to\beta\in\C$, the bulk equation
$r_{m}\beta^{2}-(E-\varepsilon_{m})\beta+l_{m}=0$ has two roots whose product is $l_{m}/r_{m}$;
for a long open chain, the GBZ condition is that these two roots be
equimodular~\cite{Yao2018,Yokomizo2019}, which fixes the GBZ independently of $E$. Provided
neither amplitude vanishes, the channel is an exactly solvable tridiagonal Toeplitz matrix at
every finite size~\cite{Noschese2013}: the diagonal gauge $D=\mathrm{diag}(\beta_{m}^{\,n})$
conjugates it into a symmetric chain with hopping $\sqrt{r_{m}l_{m}}$ (Supplemental Material~\cite{SM}, Sec.~S4). (If one amplitude does vanish,
the open chain degenerates into a one-way Jordan block and its spectrum collapses to
$\varepsilon_{m}$.) The GBZ of each channel is
thus the exact circle

\begin{equation}
|\beta_{m}|=\sqrt{\left|\frac{l_{m}}{r_{m}}\right|},\qquad
\kappa_{m}=\frac{1}{2a}\ln\left|\frac{r_{m}}{l_{m}}\right|,
\label{eq:gbzradius}
\end{equation}
with $\kappa_{m}$ the inverse skin length, positive for accumulation at the left and negative for
accumulation at the right, while the spectrum and right eigenvectors have closed-form expressions at every
finite $N$,
\begin{equation}
E_{m,j}=\varepsilon_{m}+2\sqrt{r_{m}l_{m}}\cos\tfrac{j\pi}{N+1},\quad
\psi^{(j)}_{n}=\beta_{m}^{\,n}\sin\tfrac{j\pi n}{N+1},
\label{eq:obc}
\end{equation}
with $\beta_{m}=\sqrt{l_{m}/r_{m}}$ and $j=1,\dots,N$.

Equation~\eqref{eq:obc} is exact rather than a large-$N$ approximation and has several direct
consequences. Within a fixed channel every eigenstate carries the same exponential envelope
$|\beta_{m}|^{\,n}=\ee^{-\kappa_{m}na}$, so all eigenstates of that channel share a single skin
length, $\xi_{m}=1/|\kappa_{m}|$: all eigenstates in the channel exhibit the same exponential skin localization, rather than only states near a particular spectral edge. The same envelope explains the
difference between PBC and OBC. With PBC, the eigenvalues trace
the loop $H_{m}(k)$ in the complex-energy plane. With OBC, that loop collapses onto
 the line segment $E\in\varepsilon_{m}+2\sqrt{r_{m}l_{m}}\,[-1,1]$, whose points are the standing-wave values of $\cos q$ allowed by a finite chain. Finally, the gauge
$D$ that turns the asymmetric chain into a symmetric one has condition number
$\ee^{(N-1)|\kappa_{m}|a}$, so a larger $|\kappa_{m}|$ shortens the skin length and simultaneously
worsens the conditioning of the similarity: different skin lengths come with correspondingly different sensitivities of the spectrum to
perturbations~\cite{TrefethenEmbree2005}.

The GBZ radii themselves satisfy $|\beta_{0}|=1$ and $|\beta_{-m}|=1/|\beta_{m}|$. The trivial channel sits on the unit circle and stays extended, while conjugate channels sit on reciprocal
radii [Fig.~\ref{fig:main}(b), inset]: they localize at the same rate but in opposite directions, and so accumulate at opposite ends. This pairing is an exact spatial identity. The inversion $\mathcal{P}$, defined by
$(\mathcal{P}\psi)_{n}=\psi_{N+1-n}$, exchanges sub- and superdiagonals.
Reciprocity gives $r_{-m}=l_{m}$, $l_{-m}=r_{m}$ and $\varepsilon_{-m}=\varepsilon_{m}$, so that

\begin{equation}
H_{-m}\;=\;\mathcal{P}\,H_{m}\,\mathcal{P}
\label{eq:mirror}
\end{equation}
holds exactly at every finite $N$. Conjugate partners are therefore \emph{exactly} isospectral
under OBC while carrying opposite point-gap winding numbers, and their eigenmodes are spatial mirror
images.

In the uniform chain with conserved $J_z$, \emph{reciprocal skin effect} denotes the pairing of
skin channels at opposite ends. Equation~\eqref{eq:recip} relates channels $m$ and $-m$, with
opposite windings and reciprocal GBZ radii. It does not imply spatial overlap of the partner
modes or equal total signals at the two ends. \emph{Edge balance} instead denotes the mirror
symmetry of the equal-weight summed partner density. It follows from the additional combined
inversion $\mathcal S=\mathcal P\otimes U$, expressed channel by channel in
Eq.~\eqref{eq:mirror}. A measurement has this edge balance when the excitation and detection
also respect the combined inversion. Reciprocal, $J_z$-conserving nonuniform perturbations can break this
inversion while preserving the reciprocal similarity factors inherited from uniform hopping;
they need not preserve either edge balance or the physical localization length
(Sec.~\ref{sec:stability}).

The exact OBC solution gives the real-space
manifestation of the effect; the periodic point-gap winding gives its topological label.

\section{Rotation-resolved topology: \texorpdfstring{$\mathbb{Z}$ versus $\mathbb{Z}_{2}$}{Z versus Z2}}
\label{sec:topology}

Each angular-momentum channel is an
independent directional chain, and its point-gap topology is characterized by the winding of
its spectrum about
 a reference energy $E_{B}$,

\begin{equation}
\nu_{m}(E_{B})=\frac{1}{2\pi\ii}\oint_{0}^{2\pi}\!dq\;\partial_{q}\ln\big[H_{m}(k)-E_{B}\big],
\quad q=ka.
\label{eq:winding}
\end{equation}
For the nearest-neighbor chain this equals $\pm1$ when $E_{B}$ lies inside the loop of channel
$m$, $0$ when it lies outside, and is undefined on the loop itself; every base point used below
avoids all channel spectra [Fig.~\ref{fig:main}(b)]. The complete characterization is therefore the set of channel windings
$\bm{\nu}=(\nu_{m})_{m}$, one integer per conserved channel. Channel-resolved reciprocity gives
$H_{-m}(k)=H_{m}(-k)^{\mathsf{T}}$; transposition leaves the winding unchanged while the momentum
reflection flips its sign, so $\nu_{-m}=-\nu_{m}$, and $\nu_{0}=0$ wherever the self-conjugate
channel exists---in every integer multiplet, in no half-integer one (Supplemental
Material~\cite{SM}, Sec.~S5). Microscopically the same statement reads $\kappa_{-m}=-\kappa_{m}$:
conjugate channels accumulate at opposite ends at equal rate, so no edge is preferred and
reciprocity is not violated.

The channel winding numbers contain more information than a single $\Z_{2}$ parity invariant,
and the vanishing of the total winding, $\sum_{m}\nu_{m}=0$, does not imply that every channel is topologically trivial.
That cancellation is forced by reciprocity alone. Before the channels are summed,
each is an independent one-dimensional band carrying its own integer winding, the complete
point-gap invariant of such a band~\cite{Gong2018}. For a finite multiplet, and a common base
point $E_{B}$ lying in a point gap of every channel, the symmetry-resolved phases are then
exactly
\begin{equation}
\{\bm{\nu}:\ \nu_{-m}=-\nu_{m}\}\;\cong\;\Z^{\,q},
\label{eq:group}
\end{equation}
specified completely by the windings of the $q$ channels with positive $m$. The counting is $q=j$
for integer spin $j$, $q=j+\tfrac{1}{2}$ for half-integer spin, and $q=\lfloor(n-1)/2\rfloor$ for
a $C_{n}$ rotor (Supplemental Material~\cite{SM}, Sec.~S5).

The winding of the full determinant does not distinguish these channel windings. Since $H$ is block
diagonal in $m$, the determinant of any rotation-symmetric truncation factorizes over the channels, so its winding equals $\sum_{m}\nu_{m}$ and vanishes. The zero total winding
records the cancellation between conjugate partners; it does not remove the nonzero integers
carried by the individual channels.

\begin{figure}[!t]
\includegraphics[width=\columnwidth]{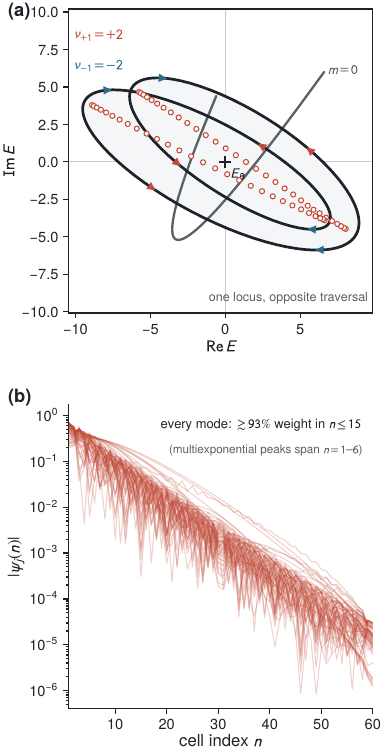}
\caption{\label{fig:range2}%
Windings beyond $\pm1$, from the range-2 rotor
[$T_{2}=t_{0}^{(2)}I+t_{1}^{(2)}\ee^{-\ii\alpha\Lz}$ with $(t_{0}^{(2)},t_{1}^{(2)})=(3,\,3\ee^{2\pi\ii/3})$ and $\alpha=2\pi/3$].
(a) Bloch locus of the $m=+1$ channel. The curve encircles the base point $E_{B}=0$ twice, so
$\nu_{+1}=+2$; the base point stays clear of every channel locus by at least $1.6$. The conjugate
partner $m=-1$ traces the same locus with the opposite orientation, $\nu_{-1}=-2$, and $m=0$
degenerates to a zero-area locus. Points: open-boundary spectrum at $N=60$, lying inside the
locus. (b) Eigenmode amplitudes of the doubly winding channel. Every mode is strongly localized to
the left, carrying $\gtrsim93\%$ of its weight in the leftmost quarter of the chain, although the
multiexponential range-2 profiles need not peak in the very first cell (the peak cells span
$n=1$--$6$).}
\end{figure}

These integers are not merely formal; they can be realized. With nearest-neighbor coupling alone, each
channel winds at most once about the loop center $E_{B}=\varepsilon_{m}$, and the sign of the
winding is simply the sign of the generator, $\nu_{m}=\sgn(g_{m})$, whenever the two amplitudes
differ in magnitude. Larger windings require only a longer coupling range. If independent rotation-diagonal amplitudes are allowed at each range
over at most $P$ cells, then for a finite multiplet every reciprocity-compatible
set of windings of magnitude at most $P$ can be realized (Appendix~\ref{app:range}). One
next-nearest coupling of the same two-path form, with $(t_{0}^{(2)},t_{1}^{(2)})=(3,\,3\ee^{2\pi\ii/3})$,
already gives $\nu_{\pm1}=\pm2$, demonstrating that the channel-resolved classification is not
restricted to a binary label
[Fig.~\ref{fig:range2}]. For the infinite $U(1)$ tower the same construction fixes the windings on
any finite window of channels. For uniformly bounded hopping, an on-site energy growing with $|m|$ supplies a sufficient regularization for a finite-support description, provided a common point gap remains open; other infinite-tower classes are outside our scope. Bulk--boundary
correspondence then holds channel by channel~\cite{Zhang2020,Okuma2020}: a nonzero winding
produces accumulation in that channel, and its sign picks the edge. The integer grading also
clarifies the role of half-integer spin, where the familiar symplectic $\Z_{2}$ NHSE appears as a
parity reduction of the conserved-channel integers.

\section{Half-integer spin: Kramers channels and the symplectic class}
\label{sec:spin}

The preceding derivation
does not require integer $m$, and for a half-integer spin-$j$ multiplet every channel-level
result---the amplitudes~\eqref{eq:rl}, the generator~\eqref{eq:generator}, the exact
solution~\eqref{eq:gbzradius}--\eqref{eq:obc}, and the mirror identity~\eqref{eq:mirror}---carries
over unchanged. Three differences arise.

The first is the symmetry class. The reversal matrix satisfies
$U^{\mathsf T}=(-1)^{2j}U$, so integer $j$ realizes class AI$^{\dagger}$ while every half-integer $j$
realizes the \emph{symplectic} class AII$^{\dagger}$; these are the two classes of transpose-type
time reversal defined by Eq.~\eqref{eq:recip}, distinguished by this transpose symmetry. The one-dimensional
point-gap classification of AII$^{\dagger}$ is $\Z_{2}$, and its hallmark is the Kramers-paired
skin effect~\cite{Kawabata2019,Okuma2020,Kawabata2020symplectic}. Here the conjugate pairs
$(m,-m)$ \emph{are} those Kramers partners: isospectral by Eq.~\eqref{eq:mirror}, of opposite
winding, with mirror-imaged modes. The second is that a half-integer multiplet has no $m=0$
channel, so for generic $\alpha$ every channel exhibits skin localization; the
exceptional angles make $m\alpha$ a multiple of $\pi$ in at least one channel.

The third concerns the counting. While $J_{z}$ is conserved, the classification
Eq.~\eqref{eq:group} and the finite-range realization of Appendix~\ref{app:range} apply without a channel whose winding is constrained to vanish, and the phases form $\Z^{\,j+1/2}$. The individual windings, however, require exact
conservation. Once arbitrary AII$^{\dagger}$-preserving mixing is admitted, the common point-gap
invariant reduces to the Pfaffian parity~\cite{Kawabata2019,Okuma2020}
\begin{equation}
\nu_{\mathrm{AII}^{\dagger}}=\sum_{m>0}\nu_{m}\ (\mathrm{mod}\ 2),
\label{eq:parity}
\end{equation}
which remains defined as long as the common point gap stays open; the derivation of this relation is given in Supplemental Material~\cite{SM}, Sec.~S8. Spin-$\tfrac{1}{2}$, with a
single winding equal to one, is therefore protected by reciprocity alone; spin-$\tfrac{3}{2}$ with
windings $(1,1)$ has a trivial $\Z_{2}$ index and yet a full graded skin effect while $J_{z}$ is
conserved; and spin-$\tfrac{5}{2}$ [Fig.~\ref{ext:sm:fig:maps}(b)] again has a nontrivial $\Z_{2}$ index.

The $\Z^q$ grading requires conservation of $J_z$ in addition to transpose-type reciprocity
(SM Sec.~\ref{ext:sm:classes}). Once a reciprocal perturbation mixes the channels, their separate
windings are no longer defined. In class AI$^{\dagger}$ the one-dimensional point-gap
classification is trivial; the studied spin-1 example loses its size-independent skin
localization length in the thermodynamic limit without closing the point gap
(Sec.~\ref{sec:stability} and Appendix~\ref{app:numerics}). Weak mixing can nevertheless leave
strong boundary accumulation in finite chains. In class AII$^{\dagger}$, the remaining invariant is the parity in Eq.~\eqref{eq:parity}: odd parity
cannot be removed by AII$^{\dagger}$-preserving mixing while the common point gap remains open,
although the individual channel windings and skin lengths are no longer protected.

This relation is analogous to the quantum spin Hall effect: when spin is conserved, the two spin subspaces have opposite Chern numbers,
and the parity of either Chern number gives the time-reversal $\Z_{2}$ invariant~\cite{Sheng2006,Prodan2009}.

The classification specifies which channel windings are allowed; a physical
implementation must also show that the required non-collinearity is compatible with reciprocity and
passivity.

\section{Passive tight-binding realization}
\label{sec:passive}

The non-collinearity need not be engineered into the hopping, because loss supplies it
automatically. The passive
realization uses identical multimode resonators---microring orbital modes or degenerate cavity
modes---linked by a Hermitian phase-carrying hopping and by
 a rotation-blind coupling running through a lossy element. One amplitude is then real and the other purely imaginary, giving
a relative phase of $\pi/2$. For
 fixed magnitudes and
rotation angle, this maximizes the channel imbalance. Uniform on-site loss of at least twice
the strength of the lossy coupling then makes each finite open chain strictly passive, through an exact operator bound.
The lossy coupling itself is not an ad
hoc matrix element: one rotation-blind Markovian loss channel per bond---physically, a
strongly
damped auxiliary resonator shared by two neighboring cells---generates it exactly, together with
the threshold on-site loss, as its
no-jump generator (Appendix~\ref{app:passive}).

Uniform loss shifts the spectrum rigidly and leaves the GBZ radii untouched; passivity and the
graded skin effect therefore coexist, with each localization edge fixed by the rotation angle
alone. The winding is likewise unchanged about a reference energy shifted by the same amount. A driven
signal nevertheless decays during propagation and may fall below the detection floor before
reaching a boundary. Observable channel-resolved accumulation therefore requires sufficient
spatial extent to resolve the skin envelope and a detectable boundary signal, with excitation
and detection that distinguish the channels. SM Sec.~\ref{ext:sm:response} gives the separate conditions
for directional contrast and source-to-boundary transmission, including their dependence on
absorption, propagation distance, and detector dynamic range.

The role of numerics is therefore to verify the exact channel predictions and to test what remains when the protecting
channel conservation is relaxed.

\section{Numerical demonstration}
\label{sec:numerics}

Exact diagonalization confirms the channel predictions: the opposite windings at $\alpha=2\pi/3$ and the collapse of
the PBC loops to the OBC spectra [Fig.~\ref{fig:main}(b)]; the graded rates $\kappa_{1}a\simeq0.17$ and $\kappa_{2}a\simeq0.83$ at $\alpha=2\pi/5$; the six Kramers-paired
 spin-$\tfrac{5}{2}$ channels; the range-2 windings of magnitude two
[Fig.~\ref{fig:range2}]; the
passivity bound; a collinear control chain showing no accumulation; and the critical, scale-free crossover that sets in once the spin-1 channels are mixed (Appendix~\ref{app:numerics}). These analytical and numerical checks place the mechanism in relation to the existing NHSE routes.

\section{Discussion}
\label{sec:discussion}

Internal rotation is the microscopic mechanism that produces the imbalance
here, rather than a hopping asymmetry introduced explicitly. The unequal channel amplitudes arise from the internal phase factors within a
coupling matrix that is itself reciprocal, which separates the mechanism from several nearby
classes of models. In Hatano--Nelson chains and their extensions the hoppings are
unequal before any symmetry resolution is performed~\cite{Hatano1996,HatanoNelson1997}; here the
Hamiltonian satisfies the reciprocity constraint~\eqref{eq:recip}, and directionality appears only
once a channel of fixed angular momentum is resolved. In reciprocal $\Z_{2}$ skin effects and related systems with internal-symmetry constraints,
spin degrees of freedom, or gauge fields~\cite{Hofmann2020,Okuma2020,Kawabata2020symplectic,Wang2024symmetry,Liu2025Z2acoustic,Sanahal2025spin,Zhang2025SU2,Han2026helical,Tang2026nonabelian}, partners related by reciprocity or with opposite pseudospins can
likewise localize at opposite ends. A single conjugate pair has opposite-edge partners, whereas higher internal
multiplets support multiple independent integer windings that cannot be represented by a single $\Z_{2}$
label [Eq.~\eqref{eq:group}]. The half-integer case
realizes class AII$^{\dagger}$, whose $\Z_{2}$ index is recovered as the parity of the grading.
Related crystalline, orbital, and noncommutative gradings are compared in Appendix~\ref{app:related}.

\section{Stability: conserved grading versus critical mixing}
\label{sec:stability}

The comparison above points to the
central stability question: the integer assigned to each channel is protected only while the channels stay decoupled. Any perturbation preserving the block
structure and keeping a point
gap open in every channel leaves all of them unchanged. Once
conjugate channels are mixed, their individual windings are no longer separately defined. For the
spatially uniform, reciprocity-preserving perturbation
$\lambda J_{x}^{2}$ studied here, any fixed $\lambda\neq0$ couples the unperturbed partners localized at opposite ends of
the spin-1 (AI$^{\dagger}$) rotor chain. In the thermodynamic limit the mixed pair has a unit-circle
GBZ and no size-independent exponential skin localization. This does not imply the immediate
disappearance of boundary accumulation in a finite chain. At weak mixing, the modes remain
strongly boundary concentrated below a mixing-dependent crossover size; longer chains
progressively approach scale-free profiles whose extent grows with $N$, with a mean inverse
participation ratio proportional to $N^{-1}$~\cite{Li2020critical,Yokomizo2021scaling}.
SM Sec.~\ref{ext:sm:norot} gives the weak-mixing data and the crossover estimate. For half-integer spin in class
AII$^{\dagger}$ the mixing likewise removes the separate integers, but the parity of Eq.~\eqref{eq:parity}
survives while the common point gap stays open.

A cell-dependent on-site potential or loss that is equal on all internal components preserves
$J_z$ and reciprocity, but generally breaks $\mathcal S=\mathcal P\otimes U$. With uniform
hopping, the conjugate eigenmodes remain
$\psi_n^{(\pm m)}=\beta_m^{\,\pm n}\phi_n$, where $\phi_n$ solves the same symmetrized-chain
problem for both partners before their separate normalization. The factors $\beta_m$ and
$\beta_m^{-1}$ are unchanged by this on-site perturbation, and the partner eigenvalues remain
identical. A generic nonuniform chain has no translational GBZ, however, so these are reciprocal
similarity factors rather than global GBZ radii. The partner densities need not be mirror
images. Moreover, disorder or a potential gradient can localize $\phi_n$ inside the sample and
change or suppress end accumulation despite the unchanged factors. For reciprocal,
$J_z$-conserving bond disorder, the factors themselves become products of the local hopping
ratios rather than powers of a fixed $\beta_m$. SM Sec.~\ref{ext:sm:nonuniform} gives both cases.

\section{Conclusion and outlook}
\label{sec:conclusion}

With this distinction between conserved grading and mixed-channel criticality in place, the
mechanism has platform-independent implementation requirements: an internal space with a nontrivial
rotation representation and two coupling paths whose complex amplitudes are not collinear. Because
only the rotation eigenphases enter the channel hopping, the internal rotation need not be a literal
mechanical rotation. It may be any conserved Abelian $U(1)$ or $C_n$ internal symmetry whose
eigenvalue supplies the link phase; no angle operator or noncommuting internal structure is required.
Accordingly, the same construction can describe orbital modes of a ring, cyclic resonator
multiplets, integer-spin manifolds, or genuine half-integer spins. In the last case
$U^{\mathsf T}=-U$ and the chain realizes class AII$^{\dagger}$ with Kramers-paired skin channels. Ring
resonators with orbital-angular-momentum modes, degenerate microwave or acoustic cavities,
coupled-waveguide arrays, and topolectrical
circuits~\cite{Scheibner2020,Ghatak2020,Brandenbourger2019,Weidemann2020,Zhang2021acoustic,Hofmann2020,Helbig2020,Wanjura2020}
are possible platforms, while genuine spin multiplets---electronic and nuclear spins such as the
$I=\tfrac{5}{2}$ of $^{173}$Yb, or cold-atom hyperfine manifolds---provide direct realizations of
the half-integer case~\cite{CazalillaRey2014}. The angular-momentum resolution enables
angular-momentum-selective boundary localization. The channel-dependent skin length also makes the spectral sensitivity channel dependent; it grows exponentially with system
size and is tunable channel by channel (any metrological claim would, of course, require a full response--noise analysis). In
higher dimensions, separable rotation-modulated couplings give a skin rate in each direction, and different channels can localize at different corners (Appendix~\ref{app:related}). Natural extensions include non-Abelian
internal rotations, line-gap symmetry constraints~\cite{Kawabata2019,Zhou2019}, and
rotation-preserving disorder~\cite{Okuma2021quadratic}. Overall, internal rotation provides a systematic route to
reciprocal skin localization with channel-resolved integer windings, even though the full system
has no preferred direction.

\begin{acknowledgments}
The authors thank Z.~Q. Zhang, Changhao Meng, and Yixin Xiao for useful discussions. This work
was supported by the Hong Kong Research Grants Council
(Grant No.~JLFS/P-603/24).
\end{acknowledgments}

\appendix

\section{Finite-range realization}
\label{app:range}

At coupling range $P$ the symbol $H_{m}(k)$ of
each channel, allowing independent rotation-diagonal amplitudes at each range, is an arbitrary Laurent polynomial of degree $P$ in $\ee^{\ii ka}$, with reciprocity
fixing the conjugate channel as its momentum reflection, $H_{-m}(k)=H_{m}(-k)^{\mathsf{T}}$.
Standard banded-Toeplitz and Laurent-symbol results~\cite{BottcherGrudsky2005}, together with
Rouch\'e's theorem, guarantee that a symbol of any winding $\nu$ with $|\nu|\le P$ exists, so for
a finite multiplet couplings of range at most $P$ attain exactly
\begin{equation}
\{\bm{\nu}\ \text{at range}\ P\}
=\{\bm{\nu}:\nu_{-m}=-\nu_{m},\,|\nu_{m}|\le P\},
\label{eq:realized}
\end{equation}
which exhausts the classifying group as $P\to\infty$. For the infinite $U(1)$ tower the same
construction prescribes the windings on any finite retained window of channels. A confining on-site spectrum, $|\varepsilon_m|\to\infty$, with uniformly bounded hopping and a positive common operator gap yields a finite-support description. Confinement is sufficient, not necessary, for a uniform gap; other infinite-tower operator classes are not classified here. In the semi-infinite Toeplitz problem $|\nu_{m}|$ is the magnitude
of the Fredholm index of the channel symbol at $E_{B}$~\cite{BottcherGrudsky2005}, and the mirror
identity~\eqref{eq:mirror} pairs equal indices at opposite edges. Figure~\ref{fig:range2} shows
the case $|\nu|=2$.

\section{Details of the passive realization}
\label{app:passive}

Non-collinearity need not be imposed on
the hopping; in a tight-binding lattice, loss supplies it. The realization uses identical multimode resonators
with internal modes labeled by $m$---clockwise and counterclockwise or orbital-angular-momentum
modes of a microring, or degenerate cavity modes---linked by a Hermitian hopping
$t_{1}=t$ that carries the internal phase factor $\ee^{-\ii\alpha\Lz}$, together with a rotation-blind coupling
$t_{0}=-\ii c$ through a lossy element (we take $t,c,\gamma_{0}\ge0$ throughout; for $c<0$ replace
$c$ by $|c|$). Channel by channel,
\begin{equation}
H_{m}(k)=\varepsilon_{0}+2t\cos(ka-m\alpha)-\ii\big(\gamma_{0}+2c\cos ka\big),
\label{eq:HTB}
\end{equation}
which superposes a real reactive band carrying the internal-rotation phase factor, an imaginary loss band that
does not, and a uniform on-site loss $\gamma_{0}$. The two amplitudes are as non-collinear as they
can be---one
real, one imaginary---so that $\im(t_{0}^{*}t_{1})=tc$ and, by Eq.~\eqref{eq:generator}, each
channel inherits $g_{m}=4tc\,\sin(m\alpha)$: loss supplies exactly the phase misalignment of
$\pi/2$ that the abstract model needs.

Passivity, $\im E\le0$, is an operator statement with an exact form. The anti-Hermitian part of
the open chain is $(H-H^{\dagger})/2\ii=-\gamma_{0}I_{N}-c\,(S+S^{\dagger})$, with $S$ the
open-chain shift, and its spectrum is $-\gamma_{0}-2c\cos[j\pi/(N+1)]$. The field-of-values
(Bendixson) inequality~\cite{HornJohnson1991}, $\lambda_{\min}(K)\le\im E\le\lambda_{\max}(K)$
with $K=(H-H^{\dagger})/(2\ii)$, therefore gives
\begin{equation}
\im E\;\le\;-\Big(\gamma_{0}-2c\cos\tfrac{\pi}{N+1}\Big).
\label{eq:passivity}
\end{equation}
A given finite open chain is dissipative when $\gamma_{0}\ge2c\cos[\pi/(N+1)]$. The
size-independent condition $\gamma_{0}\ge2c$ is necessary and sufficient under PBC
and in the thermodynamic limit, and it makes every finite open chain strictly passive.

The dissipative coupling need not be postulated as a matrix element, since a completely positive
loss channel generates it. With $a_{n,\mu}$ the amplitude of internal component $\mu$ in cell
$n$, place one rotation-blind Markovian jump on every bond and component,
\begin{equation}
L^{\mathrm{bond}}_{n,\mu}=\sqrt{2c}\,\big(a_{n,\mu}+a_{n+1,\mu}\big),
\label{eq:bondjump}
\end{equation}
alongside the coherent phase-carrying hopping $t\ee^{-\ii\alpha\Lz}$. The no-jump (drift) generator
\begin{equation}
H_{\mathrm{eff}}=H_{\mathrm{reactive}}-\frac{\ii}{2}\sum_{n,\mu}
\big(L^{\mathrm{bond}}_{n,\mu}\big)^{\dagger}L^{\mathrm{bond}}_{n,\mu}
\label{eq:heff}
\end{equation}
expands, on a periodic chain, into exactly the reciprocal dissipative hopping $t_{0}=-\ii c$
together with a uniform on-site loss $2c$, reproducing Eq.~\eqref{eq:HTB} at the passivity threshold [verified to machine precision]. Identical end jumps $\sqrt{2c}\,a_{1,\mu}$ and
$\sqrt{2c}\,a_{N,\mu}$ restore the uniform diagonal of the open chain, and extra local jumps give
any larger $\gamma_{0}$. Since the whole anti-Hermitian part is $-\ii\sum L^{\dagger}L/2$, it is
negative semidefinite and passivity holds by construction. Experimentally, one can couple each
neighboring pair equally to a single strongly damped auxiliary resonator; eliminating it
adiabatically produces the bond jump~\eqref{eq:bondjump} with $c=2g^{2}/\Gamma$, while a separate
coherent path carries the rotation.

For the classical coupled-mode realization, we use energy-normalized amplitudes, so the modal
norm balance describes energy dissipation. Each shared loss channel responds to the sum of its
two neighboring fields: in-phase contributions reinforce, whereas equal and opposite contributions
cancel. This cancellation suppresses only that bond-loss channel, not excess local absorption or
the end jumps. The loss rate in Eq.~\eqref{eq:HTB} is independent of $m$ at fixed $k$; the rotation
changes the group velocity associated with the least-damped phase pattern.

\section{Numerical checks and stability}
\label{app:numerics}

Exact diagonalization of finite
chains confirms the construction. At $\alpha=2\pi/3$ the $m=\pm1$ spectra trace one loop with
opposite orientation about the common base point $E_{B}=0.2\ii$, giving windings $\pm1$, while the
$m=0$ spectrum reduces to a zero-area segment through the origin that the base point avoids
[Fig.~\ref{fig:main}(b)]. With OBC, each loop collapses onto an interior segment. The
mirror identity~\eqref{eq:mirror} holds to machine precision, and the closed-form
eigenvectors~\eqref{eq:obc} satisfy the eigenvalue equation to a residual of $10^{-15}$.

Because $C_{3}$ folding leaves only a single rate ($\kappa_{3}=0$ and
$|\kappa_{1}|=|\kappa_{2}|$), the grading of localization \emph{lengths} is shown instead at
$\alpha=2\pi/5$ [Fig.~\ref{ext:sm:fig:maps}(a)]. There
$m\equiv1,2\ (\mathrm{mod}\ 5)$ accumulate at the
left with $\kappa_{1}a\simeq0.17$ and $\kappa_{2}a\simeq0.83$, their partners $m\equiv4,3$ at the
right at the mirrored rates, and $m\equiv0$ stays extended---four distinct skin channels that
cannot be represented by a single $\Z_{2}$ label, with the rates given by
Eq.~\eqref{eq:gbzradius} for the channels shown.

The range-2 rotor of the main text gives windings $(-2,0,+2)$ about a gapped base point at the
origin, with every mode of the plotted $m=+1$ channel strongly localized to the left [Fig.~\ref{fig:range2}]. The anti-Hermitian
spectrum behind Eq.~\eqref{eq:passivity} matches its closed form to $10^{-16}$, and a collinear
control chain with real $t_{1}$ gives real spectra and no accumulation, exactly as the vanishing
of the generator requires (Figs.~S1, S2, S4--S6). For
spin-$\tfrac{5}{2}$ at $\alpha=2\pi/7$ the reciprocity constraint~\eqref{eq:recip} holds exactly
with $U^{\mathsf T}=-U$, all six channels localize at three distinct rates
[Fig.~\ref{ext:sm:fig:maps}(b)], and the parities of spin-$\tfrac{1}{2}$, $\tfrac{3}{2}$ and $\tfrac{5}{2}$ are $1$, $0$ and $1$.

Internal rotation generates the imbalance and defines the channel grading, but exact conservation
protects the individual integers. We test channel mixing in the integer (AI$^{\dagger}$) spin-1 rotor with a spatially uniform, reciprocity-preserving on-site term $\lambda J_{x}^{2}$, which spoils
$[H,J_{z}]=0$ while keeping both the reciprocity constraint and the inversion symmetry
$\mathcal{S}$ exact. The mixed $\pm1$ pair has a palindromic non-Bloch quartic, so its GBZ roots
come in reciprocal pairs and the two middle roots lie on the unit circle: the thermodynamic skin
length diverges for every fixed $\lambda\neq0$~\cite{Li2020critical,Yokomizo2021scaling}.
This result does not exclude strong boundary concentration at weak mixing in a finite chain
(SM Sec.~\ref{ext:sm:norot}, Table~\ref{ext:sm:tab:mixscaling}(b)). Numerically at
$\lambda=1$, averaging over all three channels gives $N\langle\mathrm{IPR}\rangle=1.23$--$1.29$
for $N=12$--$64$, so the localization
length grows in proportion to $N$; the open-boundary spectrum converges to the periodic one with a
Hausdorff distance falling as $1/N$; and the common point gap stays open along the whole
interpolation, $\min_{\lambda\in[0,1],k}\sigma_{\min}[H_{\lambda}(k)-0.2\ii]\simeq0.195$. The
size-independent graded skin localization is therefore lost in the thermodynamic limit
\emph{without} any gap closing, and the bulk weight of $0.44$
at $N=48$ is a finite-size critical profile rather than an intensive skin effect. Accumulation
nonetheless remains exactly edge balanced, to $10^{-14}$, through $\mathcal{S}$, an extra symmetry of the
uniform chain that reciprocity alone does not imply. Only the odd AII$^{\dagger}$ parity survives
arbitrary reciprocity-preserving mixing.

\section{Relation to other graded skin effects}
\label{app:related}

Spatial crystalline gradings beyond
$\Z_{2}$---glide-protected $\Z_{4}$~\cite{Ishikawa2024glide} or $C_{n}$ symmetry-indicator
classifications~\cite{WangBenalcazar2025}---are tied to lattice symmetries, whereas the grading
found here is controlled by an \emph{internal} rotation and grows without bound with the internal
multiplet. General symmetry classifications systematically organize intrinsic point-gap phases across
non-Hermitian symmetry classes~\cite{Shiozaki2026intrinsic}, but do not by themselves specify the
microscopic mechanism that generates the channel hierarchy considered here. Sector-dependent skin
localization can also arise from Hilbert-space fragmentation in interacting systems~\cite{WangLi2025fragmented},
while symmetry-fractionalized non-Hermitian Luttinger liquids exhibit an emergent decoupling of distinct
symmetry sectors~\cite{Ekman2026fractionalized}; both are interaction-driven settings rather than a
single-particle channel hierarchy generated by conserved internal rotation. Orbital acoustic and OAM-polariton systems~\cite{Chen2025orbital,Xu2021ring} use implementation-specific orbital degrees of freedom, while optical circulation can also produce an OAM output~\cite{Takeda2025circulating}; here the internal rotation eigenvalue provides
a common channel label, and the conserved channels yield the integer grading across the multiplet. Finally, in contrast to
the spatial-noncommutativity route~\cite{WeiKou2026}, no noncommuting structure enters at all:
the mechanism is an Abelian rotation phase weighting two interfering reciprocal paths, with the
skin direction set by that phase and reversed by $m\to-m$. In higher dimensions, separable
rotation-modulated couplings carry one generator per direction, proportional to
$\sin(m\alpha_{a})$, which sends different channels to different corners while the equal-weight partner-mode density occupies antipodal corners---linking the mechanism to higher-order and geometry-dependent skin
effects~\cite{Kawabata2020higher,Zhang2021acoustic,Zhang2022higherD}.

\nocite{Longhi2019,Song2019,Xue2021,Varshalovich1988}
\bibliography{refs}

\end{document}